\documentclass[12pt]{article}
\usepackage{epsfig}

\usepackage{microtype}
\usepackage{lineno}  
\usepackage{xspace} 
\usepackage{ifthen} 
\usepackage{graphicx}  
\newboolean{articletitles}
\setboolean{articletitles}{false} 
\newboolean{uprightparticles}
\setboolean{uprightparticles}{false} 

\def\ux85 {\mbox{UX85}\xspace}

\ifthenelse{\boolean{uprightparticles}}%
{

 \def\Pmu         {\ensuremath{\upmu}\xspace}

 \def\Ppi         {\ensuremath{\uppi}\xspace}

 \def\Ptau        {\ensuremath{\uptau}\xspace}

 \def\Ppsi        {\ensuremath{\uppsi}\xspace}

 \def\PDelta      {\ensuremath{\Delta}\xspace}                 
 \def\PXi      {\ensuremath{\Xi}\xspace}                 
 \def\PLambda      {\ensuremath{\Lambda}\xspace}                 
 \def\PSigma      {\ensuremath{\Sigma}\xspace}                 
 \def\POmega      {\ensuremath{\Omega}\xspace}                 
 \def\PUpsilon      {\ensuremath{\Upsilon}\xspace}                 
 
 \def\PB      {\ensuremath{\mathrm{B}}\xspace}                 
                  
 \def\PD      {\ensuremath{\mathrm{D}}\xspace}

 \def\PJ      {\ensuremath{\mathrm{J}}\xspace}                 
 \def\PK      {\ensuremath{\mathrm{K}}\xspace}

 \def\Pe      {\ensuremath{\mathrm{e}}\xspace}

 \def\Pi      {\ensuremath{\mathrm{i}}\xspace}

 \def\Ps      {\ensuremath{\mathrm{s}}\xspace}

}
{

 \def\Pmu         {\ensuremath{\mu}\xspace}

 \def\Ppi         {\ensuremath{\pi}\xspace}

 \def\Ptau        {\ensuremath{\tau}\xspace}

 \def\Ppsi        {\ensuremath{\psi}\xspace}                 
                  
 \mathchardef\PDelta="7101
 \mathchardef\PXi="7104
 \mathchardef\PLambda="7103
 \mathchardef\PSigma="7106
 \mathchardef\POmega="710A
 \mathchardef\PUpsilon="7107
                  
 \def\PB      {\ensuremath{B}\xspace}                 
                  
 \def\PD      {\ensuremath{D}\xspace}

 \def\PJ      {\ensuremath{J}\xspace}                 
 \def\PK      {\ensuremath{K}\xspace}

 \def\Pe      {\ensuremath{e}\xspace}

 \def\Pi      {\ensuremath{i}\xspace}

 \def\Ps      {\ensuremath{s}\xspace}

}

\def\epem       {\ensuremath{\Pe^+\Pe^-}\xspace}

\def\mup        {\ensuremath{\Pmu^+}\xspace}
\def\mumu       {\ensuremath{\Pmu^+\Pmu^-}\xspace}

\def\taup       {\ensuremath{\Ptau^+}\xspace}

\def\squark    {\ensuremath{\Ps}\xspace}

\def\pion  {\ensuremath{\Ppi}\xspace}
\def\piz   {\ensuremath{\pion^0}\xspace}

\def\pip   {\ensuremath{\pion^+}\xspace}
\def\pim   {\ensuremath{\pion^-}\xspace}

\def\kaon  {\ensuremath{\PK}\xspace}
  \def\Kbar  {\kern 0.2em\overline{\kern -0.2em \PK}{}\xspace}

\def\Kz    {\ensuremath{\kaon^0}\xspace}
\def\Kzb   {\ensuremath{\Kbar^0}\xspace}
\def\KzKzb {\ensuremath{\Kz \kern -0.16em \Kzb}\xspace}
\def\Kp    {\ensuremath{\kaon^+}\xspace}
\def\Km    {\ensuremath{\kaon^-}\xspace}

\def\KpKm  {\ensuremath{\Kp \kern -0.16em \Km}\xspace}
\def\KS    {\ensuremath{\kaon^0_{\rm\scriptscriptstyle S}}\xspace} 
 
\def\Kstarz  {\ensuremath{\kaon^{*0}}\xspace}

  \def\Dbar    {\kern 0.2em\overline{\kern -0.2em \PD}{}\xspace}
\def\D       {\ensuremath{\PD}\xspace}

\def\Dz      {\ensuremath{\D^0}\xspace}
\def\Dzb     {\ensuremath{\Dbar^0}\xspace}
\def\DzDzb   {\ensuremath{\Dz {\kern -0.16em \Dzb}}\xspace}
\def\Dp      {\ensuremath{\D^+}\xspace}
\def\Dm      {\ensuremath{\D^-}\xspace}

\def\DpDm    {\ensuremath{\Dp {\kern -0.16em \Dm}}\xspace}

\def\Dsp     {\ensuremath{\D^+_\squark}\xspace}
\def\Dsm     {\ensuremath{\D^-_\squark}\xspace}

\def\B       {\ensuremath{\PB}\xspace}
  \def\Bbar    {\kern 0.18em\overline{\kern -0.18em \PB}{}\xspace}

\def\Bz      {\ensuremath{\B^0}\xspace}

\def\Bu      {\ensuremath{\B^+}\xspace}

\def\Bp      {\ensuremath{\Bu}\xspace}

\def\Bd      {\ensuremath{\B^0}\xspace}
\def\Bs      {\ensuremath{\B^0_\squark}\xspace}

\def\jpsi     {\ensuremath{{\PJ\mskip -3mu/\mskip -2mu\Ppsi\mskip 2mu}}\xspace}

  \def\Y#1S{\ensuremath{\PUpsilon{(#1S)}}\xspace}

\def\Lbar {\ensuremath{\kern 0.1em\overline{\kern -0.1em\PLambda}}\xspace}

\def\to                 {\ensuremath{\rightarrow}\xspace}

\def\CP                {\ensuremath{C\!P}\xspace}
\def\CPT               {\ensuremath{C\!PT}\xspace}

\def\AT#1     {\ensuremath{A_{\mathrm{T}}^{#1}}\xspace}           

\def\C#1      {\ensuremath{\mathcal{C}_{#1}}\xspace}                       
\def\Cp#1     {\ensuremath{\mathcal{C}_{#1}^{'}}\xspace}                    
\def\Ceff#1   {\ensuremath{\mathcal{C}_{#1}^{\mathrm{(eff)}}}\xspace}        
\def\Cpeff#1  {\ensuremath{\mathcal{C}_{#1}^{'\mathrm{(eff)}}}\xspace}       
\def\Ope#1    {\ensuremath{\mathcal{O}_{#1}}\xspace}                       
\def\Opep#1   {\ensuremath{\mathcal{O}_{#1}^{'}}\xspace}                    

\newcommand{\tev}{\ensuremath{\mathrm{\,Te\kern -0.1em V}}\xspace}
\newcommand{\gev}{\ensuremath{\mathrm{\,Ge\kern -0.1em V}}\xspace}
\newcommand{\mev}{\ensuremath{\mathrm{\,Me\kern -0.1em V}}\xspace}
\newcommand{\kev}{\ensuremath{\mathrm{\,ke\kern -0.1em V}}\xspace}
\newcommand{\ev}{\ensuremath{\mathrm{\,e\kern -0.1em V}}\xspace}
\newcommand{\gevc}{\ensuremath{{\mathrm{\,Ge\kern -0.1em V\!/}c}}\xspace}
\newcommand{\mevc}{\ensuremath{{\mathrm{\,Me\kern -0.1em V\!/}c}}\xspace}
\newcommand{\gevcc}{\ensuremath{{\mathrm{\,Ge\kern -0.1em V\!/}c^2}}\xspace}
\newcommand{\gevgevcccc}{\ensuremath{{\mathrm{\,Ge\kern -0.1em V^2\!/}c^4}}\xspace}
\newcommand{\mevcc}{\ensuremath{{\mathrm{\,Me\kern -0.1em V\!/}c^2}}\xspace}

\def\invfb   {\ensuremath{\mbox{\,fb}^{-1}}\xspace}

\def\gsim{{~\raise.15em\hbox{$>$}\kern-.85em
          \lower.35em\hbox{$\sim$}~}\xspace}
\def\lsim{{~\raise.15em\hbox{$<$}\kern-.85em
          \lower.35em\hbox{$\sim$}~}\xspace}

\def\tell1  {TELL1\xspace}
\def\ukl1   {UKL1\xspace}

\newcommand{\eg}{\mbox{\itshape e.g.}\xspace}
\newcommand{\ie}{\mbox{\itshape i.e.}}

\usepackage{hyperref}    
\usepackage[all]{hypcap} 

\usepackage{cite} 
\usepackage{mciteplus}

\def\Title#1{\begin{center} {\Large {\bf #1} } \end{center}}

\begin{document}

\Title{What We've Learned from Heavy Flavour Experiments Since CKM2010}

\bigskip\bigskip

\begin{raggedright}  

{\it Tim Gershon\index{Gershon, T.}\\
Department of Physics, University of Warwick, Coventry, United Kingdom \\
and \\
European Organization for Nuclear Research (CERN), Geneva, Switzerland}
\bigskip\bigskip
\end{raggedright}

\vspace{5ex}

\abstract{
  A summary of the latest experimental results in heavy flavour physics is presented.
}

\vspace{10ex}

\begin{center}
  Proceedings of CKM 2012, \\
  the $7^{\rm th}$ International Workshop on the CKM Unitarity Triangle, \\
  University of Cincinnati, USA, 28 September -- 2 October 2012.
\end{center}

\vfill


\clearpage

\section{Preliminaries}

Heavy flavour physics has seen enormous progress over the last decade or so, primarily through the successes of the ``$B$ factory'' experiments, BaBar~\cite{Aubert:2001tu} and Belle~\cite{Abashian:2000cg}, complemented by several notable achievements by the CDF~\cite{Abe:1988me} and D0~\cite{Abazov:2005pn} experiments based at the Tevatron.
Most recently the LHC experiments have taken on the mantle of producing world leading results in this field (and others); the LHCb~\cite{Alves:2008zz} experiment in particular is breaking new ground in beauty and charm physics, while ATLAS~\cite{Aad:2008zzm} and CMS~\cite{Chatrchyan:2008aa} also have notable $b$ physics components to their programmes.

It must be acknowledged that the successes of the experiments is a direct consequence of the excellent performance of the accelerators:
\begin{itemize}
\item PEP-II delivered $433 \invfb$ of $\epem$ collisions at the $\Upsilon(4S)$ resonance to BaBar, plus additional data at other energies;
\item KEK-B delivered $711 \invfb$ of $\epem$ collisions at the $\Upsilon(4S)$ resonance, and $121 \invfb$ at the $\Upsilon(5S)$ resonance, to Belle, plus additional data at other energies;
\item the Tevatron delivered $12 \invfb$ of $p\bar{p}$ collisions at $1.96 \tev$ to each of CDF and D0; 
\item during 2011, the LHC delivered $1.2$ ($6$) \invfb of $pp$ collisions at $7 \tev$ to LHCb (each of ATLAS and CMS), and by the time of CKM2012 the 2012 run of the LHC had added $1.5$ ($15$) \invfb of $8 \tev$ $pp$ collisions to these respective data samples; additional $pp$ collision data has been delivered at other energies.
\end{itemize}

In order to exploit maximally these unprecedented data samples requires novel detector and innovative analysis techniques.  
It is impossible to do justice to the wide range of developments in this area in a single sentence, but as a brief sample of important breakthroughs: the particle identification capability provided, \eg, by the BaBar DIRC detector~\cite{Aubert:2001tu}, neural network based event reconstruction as used at Belle~\cite{Feindt:2011mr}, trigger algorithms that enable heavy flavour physics measurements at hadron colliders by exploiting the displaced vertex signature~\cite{Ristori:2010zz,Aaij:2012me} and silicon detectors to precisely determine the vertex position, such as the LHCb VELO~\cite{Alves:2008zz}.

\section{\boldmath \CP violation}

\CP violation remains one of the deep mysteries of the Standard Model (SM).
Although it is elegantly accommodated within the CKM matrix~\cite{Cabibbo:1963yz,Kobayashi:1973fv}, and although all measurements to date are consistent with the sole source of \CP violation being the KM phase, continued experimental investigation is well motivated in order to try to identify additional sources.
Indeed, new sources of \CP violation must exist in nature in order to generate the baryon asymmetry of the Universe.
Table~\ref{tab:CP} lists the channels where \CP violation has been observed to date.
The list has become longer over the past few years, and only recently has the first observation of \CP violation in a charged particle decay~\cite{Aaij:2012kz} been achieved.
There is still no $5\sigma$ measurement of \CP violation in the $\Bs$ sector, or in the charm sector, or in any baryon decay, or in the lepton sector.
Clearly, there remains much to investigate.

\begin{table}[!htb]
\centering
\caption{\small
  Observations of \CP violation with greater than $5\sigma$ significance, separated by quark-level transition (modified from the review by D.~Kirkby and Y.~Nir in Ref.~\cite{Beringer:1900zz}, including averages from Ref.~\cite{Amhis:2012bh}).
  $^{(*)}$~New results presented at CKM2012 are not included.
}
\label{tab:CP}
\vspace{1ex}
\resizebox{0.99\textwidth}{!}{
\begin{tabular}{ll}
\hline
\multicolumn{2}{c}{Kaon sector} \\
\hline
& $\left| \epsilon_K \right| = (2.228 \pm 0.011) \times 10^{-3}$~\cite{Beringer:1900zz} \\
& ${\rm Re}(\epsilon^\prime /\epsilon_K) = (1.65 \pm 0.26) \times 10^{-3}$~\cite{Batley:2002gn,Abouzaid:2010ny} \\
\hline
\multicolumn{2}{c}{$b$ sector} \\
\hline
$b \to c\bar{c}s$ & $S_{\psi K^0} = +0.679 \pm 0.020$~\cite{Aubert:2009aw,Adachi:2012et}$^{(*)}$ \\
$b \to q\bar{q}s$ & $S_{\eta^\prime K^0} = +0.59 \pm 0.07$~\cite{:2008se,Chen:2006nk}, $S_{\phi K^0} = +0.74\,^{+0.11}_{-0.13}$~\cite{Lees:2012kxa,Nakahama:2010nj}, \\
& \ \ \ \ $S_{f_0(980)K^0} = +0.69\,^{+0.10}_{-0.12}$~\cite{Lees:2012kxa,Nakahama:2010nj,Aubert:2009me,Dalseno:2008wwa}, $S_{K^+K^-K^0} = +0.68\,^{+0.09}_{-0.10}$~\cite{Lees:2012kxa,Nakahama:2010nj} \\
$b \to u\bar{u}d$ & $S_{\pi^+\pi^-} = -0.65 \pm 0.07$~\cite{Lees:2012kx,Ishino:2006if}$^{(*)}$, $C_{\pi^+\pi^-} = -0.36 \pm 0.06$~\cite{Lees:2012kx,Ishino:2006if}$^{(*)}$ \\
$b \to c\bar{c}d$ & $S_{\psi \pi^0} = -0.93 \pm 0.15$~\cite{Aubert:2008bs,Lee:2007wd}, $S_{D^+D^-} = -0.98 \pm 0.17$~\cite{Aubert:2008ah,Rohrken:2012ta}, \\
& \ \ \ \ $S_{D^{*\pm}D^{\mp}} = -0.73 \pm 0.11$~\cite{Aubert:2008ah,Rohrken:2012ta}, $S_{D^{*+}D^{*-}} = -0.71 \pm 0.09$~\cite{Aubert:2008ah,Lees:2012px,Kronenbitter:2012ha} \\
$b \to u\bar{u}s$ & $A_{K^\mp\pi^\pm} = -0.087 \pm 0.008$~\cite{Lees:2012kx,Lin:2008zzaa,Aaltonen:2011qt,Aaij:2012qe} \\
$b \to c\bar{u}s/u\bar{c}s$ & $A_{D_{CP+}K^\pm} = +0.19 \pm 0.03$~\cite{delAmoSanchez:2010ji,Abe:2006hc,Aaltonen:2009hz,Aaij:2012kz} \\
\hline
\end{tabular}
}
\end{table}

\begin{figure}[!htb]
\centering
\includegraphics[width=0.40\textwidth]{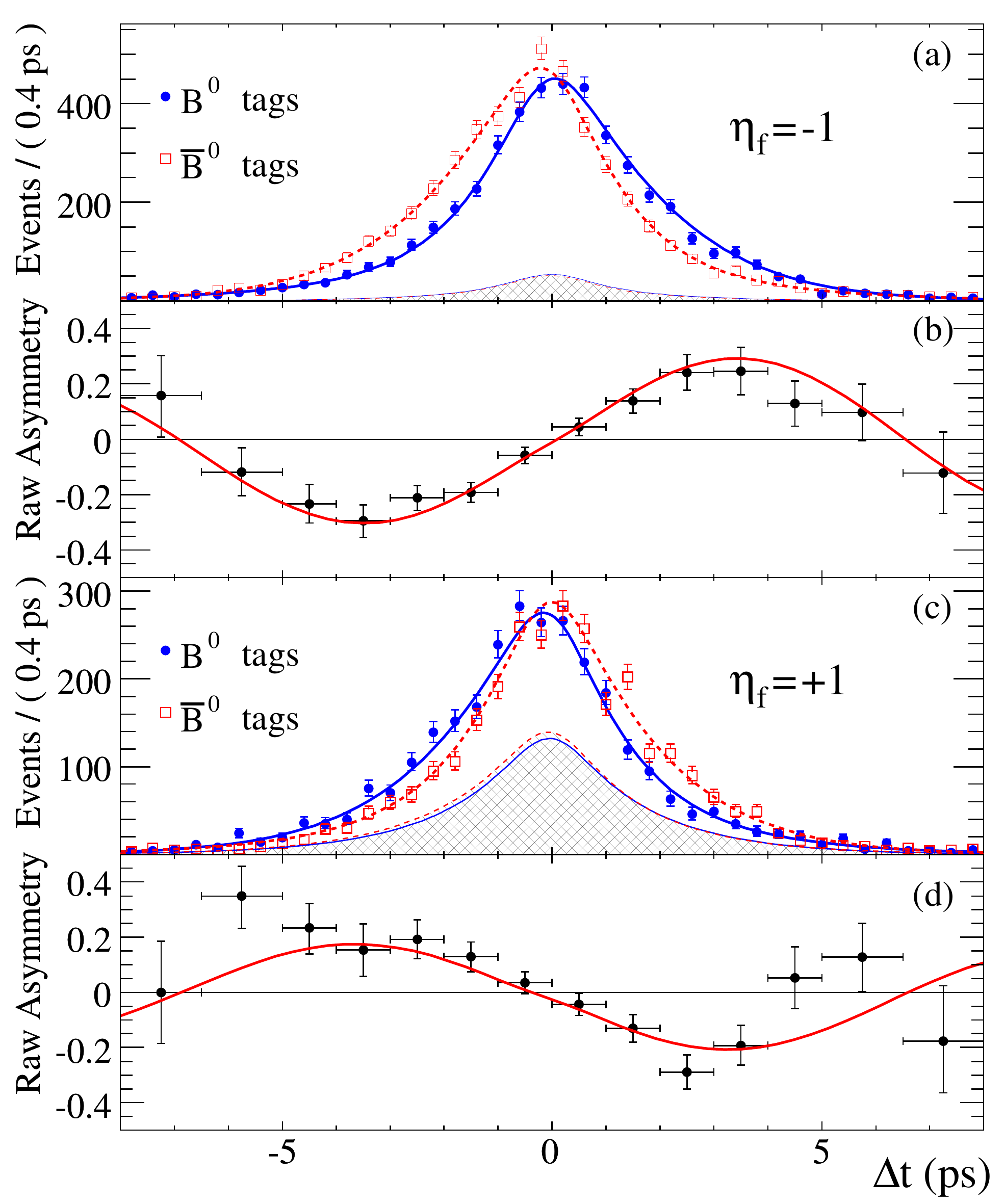}
\hspace{3mm}
\includegraphics[width=0.265\textwidth]{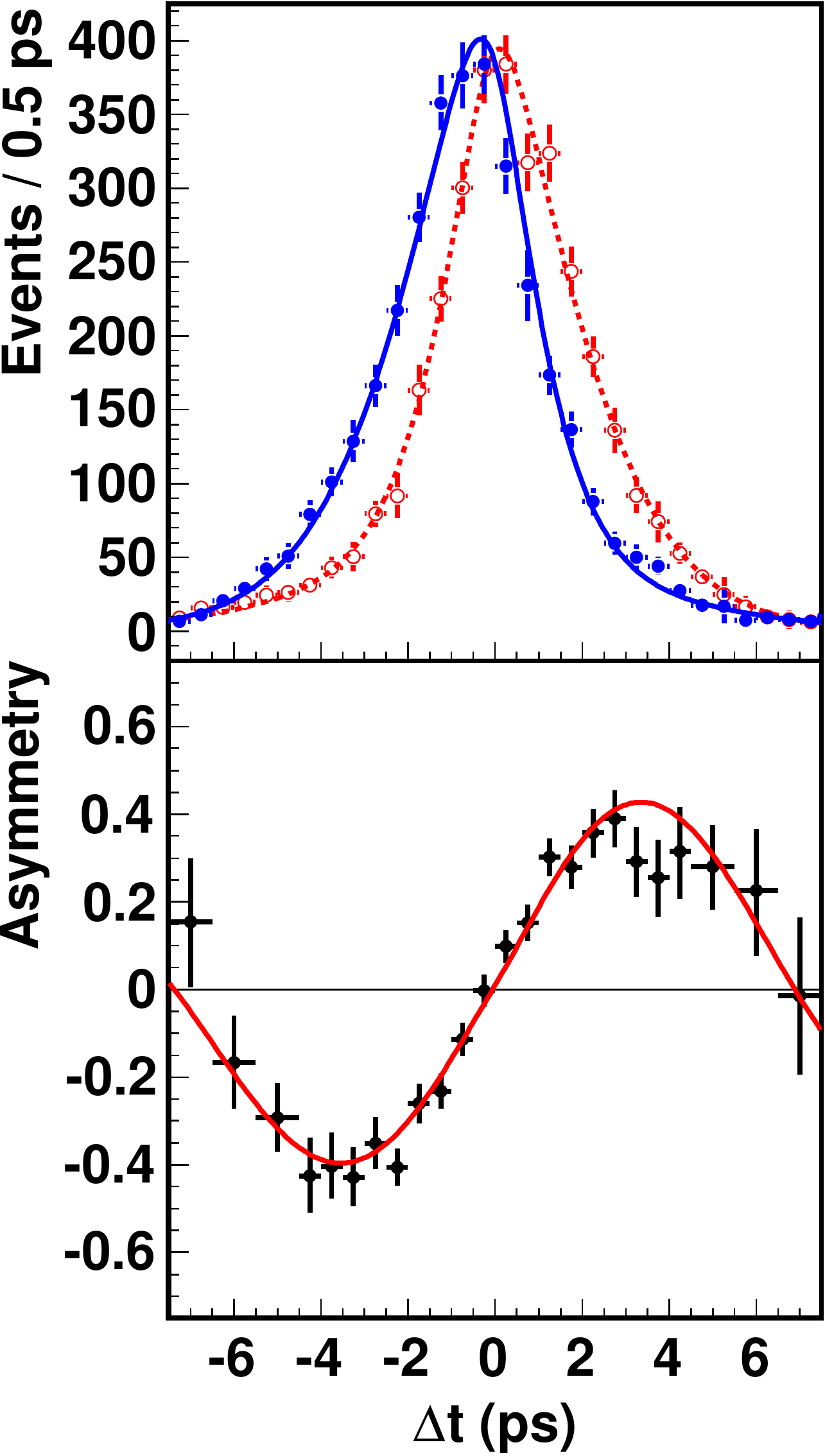}
\caption{\small
  Results from (left) BaBar~\cite{Aubert:2009aw} and (right) Belle~\cite{Adachi:2012et} on the determination of $\sin2\beta$.  
}
\label{fig:sin2beta}
\end{figure}

The determination of $\sin(2\beta)$, where $\beta$ is the angle of the CKM Unitarity Triangle, provided a spectacular confirmation of the KM scheme~\cite{Aubert:2001nu,Abe:2001xe}.
The latest (and, excluding upgrades, most likely final) results from BaBar~\cite{Aubert:2009aw} and Belle~\cite{Adachi:2012et} shown in Fig.~\ref{fig:sin2beta} give a clear visual confirmation of the large \CP violation effect.
Further improvement in precision is still needed, and therefore it is encouraging that LHCb has presented its first results on this channel~\cite{:2012ke,Wishahi}.
The quantum correlated nature of the $B$ mesons produced in $\Upsilon(4S) \to B\bar{B}$ decays can in addition be used to make novel tests of quantum mechanics~\cite{Bertlmann:1996at,Banuls:1999aj,Go:2007ww,Bernabeu:2012ab}.  
Recently, the BaBar collaboration has extended its $\sin(2\beta)$ analysis to demonstrate that $T$ is violated, as well as \CP, in the $\Bz \to \jpsi \Kz$ system (but \CPT is conserved, as expected)~\cite{:2012kn}.

Direct \CP violation has been established for several years in both kaon~\cite{Barr:1993rx,Fanti:1999nm,AlaviHarati:1999xp} and $B$~\cite{Aubert:2004qm,Chao:2004mn} sectors.
However, recently new observations have been made of very large, near maximal, direct \CP violation effects.
The LHCb collaboration has reported preliminary studies of the variation of \CP asymmetries across the Dalitz plots of $\Bp$ decays to the three-body final states $\Kp\Km\Kp$, $\Kp\pip\pim$~\cite{LHCb-CONF-2012-018}, $\Kp\Km\pip$ and $\pip\pim\pip$~\cite{LHCb-CONF-2012-028,Miranda}, revealing regions of phase space where the direct \CP violation effect appears to be in excess of $50\,\%$.
Detailed studies will be necessary to understand the origin of these effects.

Perhaps of even greater interest is the possibility of \CP violation in the charm system, which is already bounded to be small, but is (at least, na\"ively) expected to be much smaller in the SM.
However, several measurements~\cite{Aaij:2011in,Collaboration:2012qw} now point to non-zero direct \CP violation effects in the charm system, specifically in the difference in \CP violation between $\Dz \to \Kp\Km$ and $\Dz \to \pip\pim$ decays, with the world average being $\Delta a_{\CP}^{\rm dir} = (-0.68 \pm 0.15)\,\%$~\cite{Amhis:2012bh}.
The emerging consensus (reviewed in Ref.~\cite{Bediaga:2012py}) is that an effect of this size is unlikely in the SM, but could in principle be generated by QCD corrections.
Improved measurements of the \CP violation effects in several decay channels are needed to resolve the situation.
These measurements must aim at per mille level precision, and will be challenging experimentally.

\begin{figure}[!htb]
\centering
\includegraphics[width=0.4\textwidth]{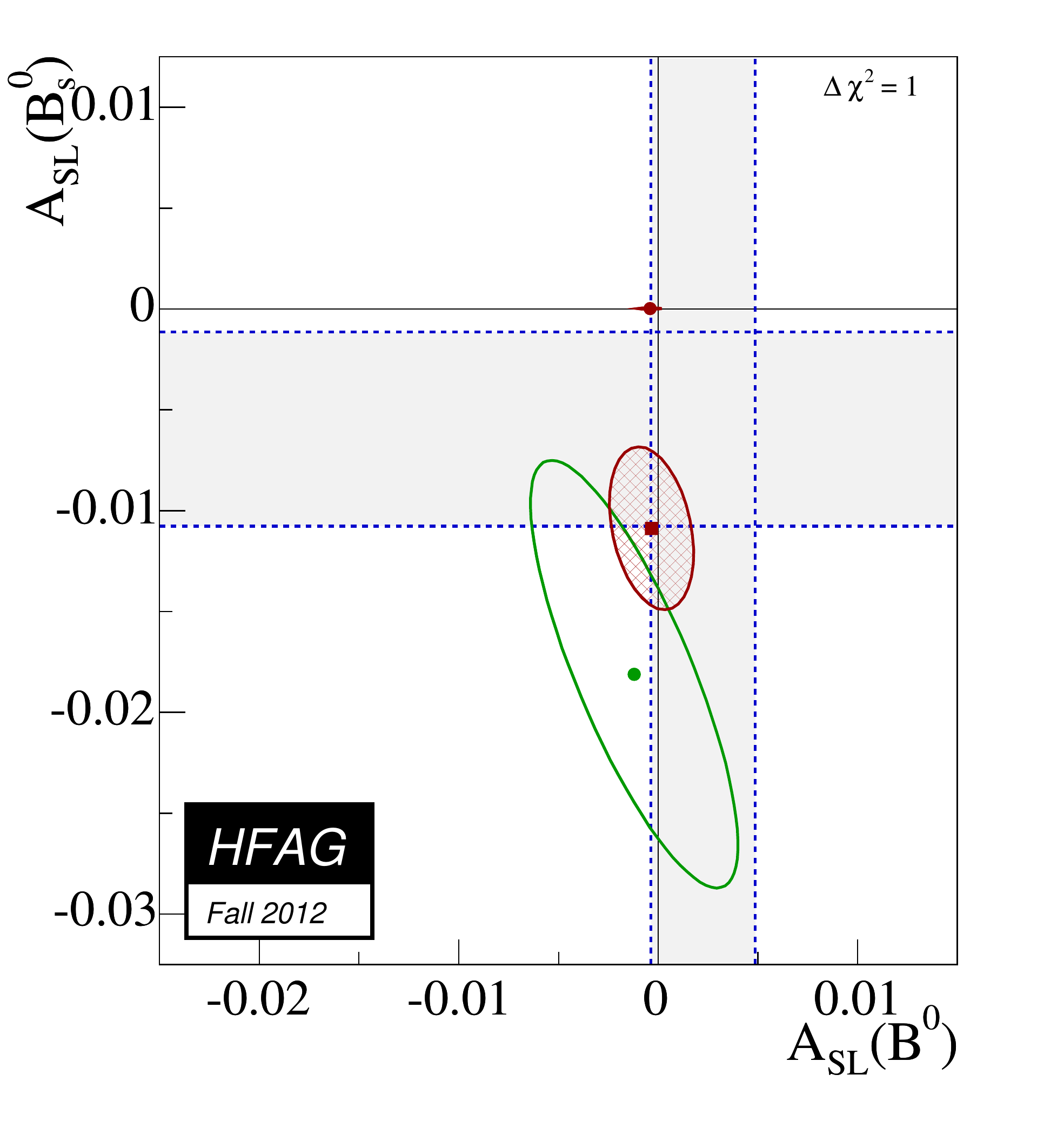}
\caption{\small
  World average of constraints on the parameters describing \CP violation in \Bd and \Bs mixing, $a_{\rm sl}^d$ and $a_{\rm sl}^s$.
  The red point close to $(0,0)$ is the SM prediction~\cite{Lenz:2011ti}.
  The green ellipse comes from the D0 inclusive same-sign dimuon analysis~\cite{Abazov:2011yk}; the blue shaded bands give the world average constraints on $a_{\rm sl}^d$ and $a_{\rm sl}^s$ individually; the red ellipse is the world average including all constraints~\cite{Amhis:2012bh}.  
}
\label{fig:qovp}
\end{figure}

A similar challenge is faced in the $B$ sector, where the possibility of anomalous \CP violation in mixing has been raised by a measurement by D0~\cite{Abazov:2011yk} of the inclusive same-sign dimuon asymmetry, which is $3.9\sigma$ away from the SM prediction (of very close to zero~\cite{Lenz:2011ti}).
This measurement is sensitive to an approximately equal combination of the parameters of \CP violation in \Bd and \Bs mixing, $a_{\rm sl}^d$ and $a_{\rm sl}^s$, however some sensitivity to the source of the asymmetry can be obtained by applying additional constraints on the impact parameter to obtain a sample enriched in either oscillated $\Bd$ or $\Bs$ candidates. 
In addition, $a_{\rm sl}^d$ and $a_{\rm sl}^s$ can be measured individually, with recent new results from D0~\cite{Abazov:2012zz,:2012uia}, LHCb~\cite{LHCb-CONF-2012-022} and BaBar~\cite{Margoni}.
The latest world average, shown in Fig.~\ref{fig:qovp}, gives
$a_{\rm sl}^d = -0.0003 \pm 0.0021$,
$a_{\rm sl}^s = -0.0109 \pm 0.0040$~\cite{Amhis:2012bh},
$2.4\sigma$ away from the SM prediction.
Improved measurements of both $a_{\rm sl}^d$ and $a_{\rm sl}^s$ are needed.

\section{The Unitary Triangle}

\begin{figure}[!htb]
\centering
\includegraphics[width=0.40\textwidth]{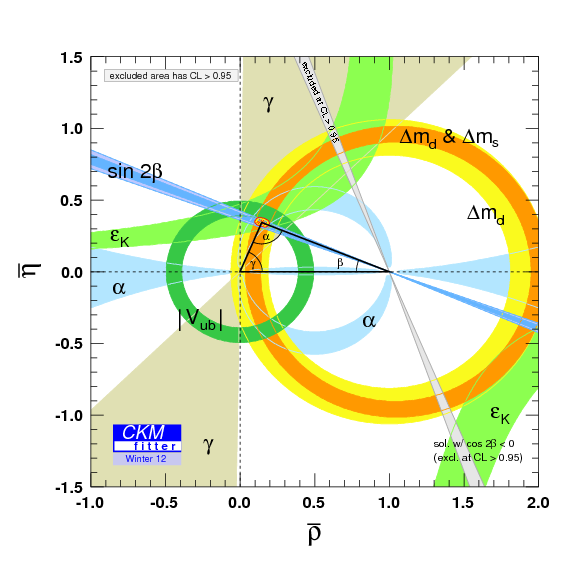}
\hspace{3mm}
\includegraphics[width=0.41\textwidth]{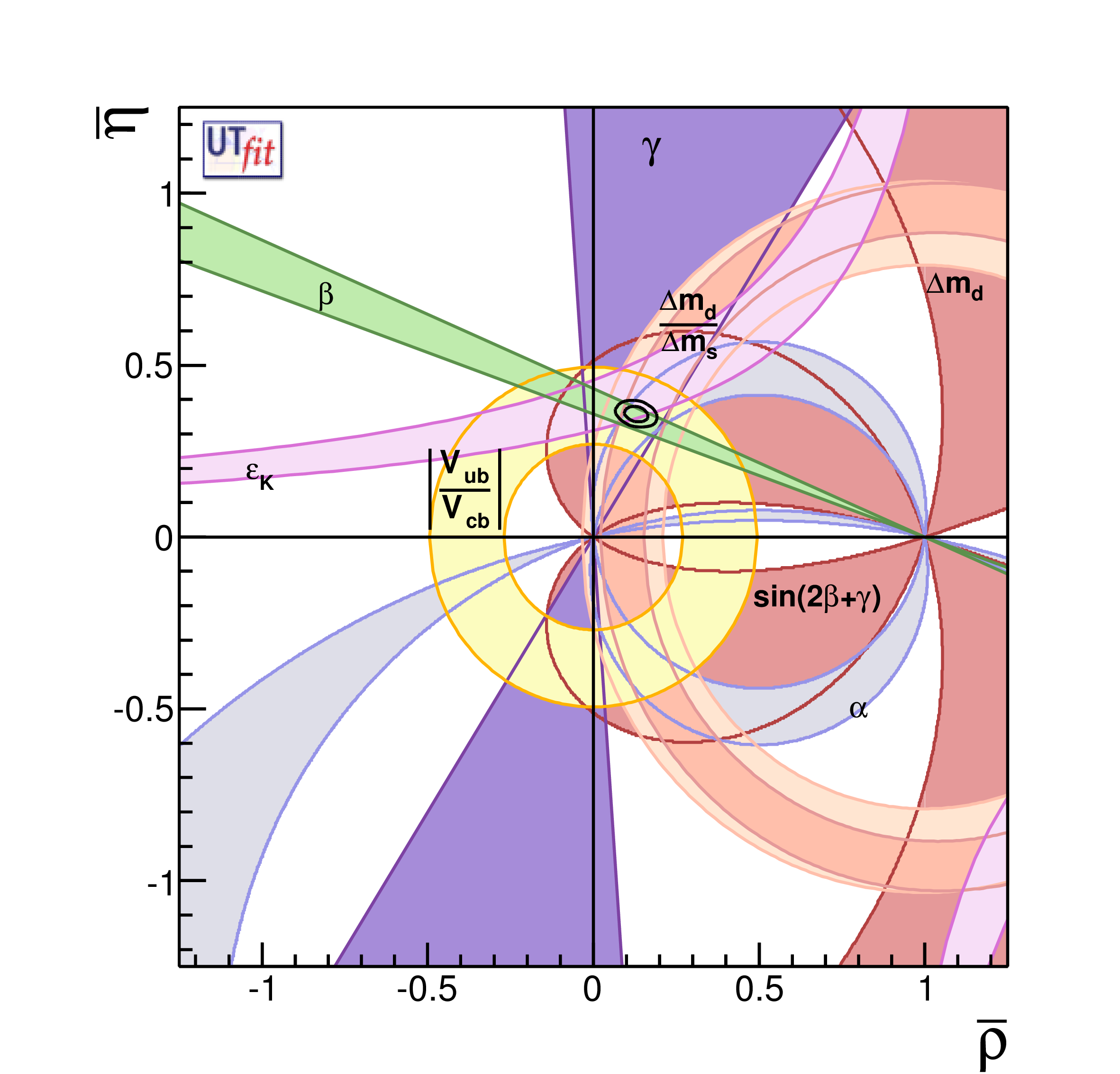}
\caption{\small
  Constraints on the Unitarity Triangle as compiled by (left) CKMfitter~\cite{Charles:2004jd}, (right) UTfit~\cite{Bona:2005vz}.
}
\label{fig:ckmfit}
\end{figure}

Many of the most important results in heavy flavour physics can be depicted as constraints in the so-called CKM Unitarity Triangle, that visualises the unitarity relation between the first and third columns of the CKM matrix.
The constraints include measurements of the angles and lengths of the sides of the triangle.\footnote{
  Here the $\alpha, \beta, \gamma$ notation is used for the angles.
}
Although other similar triangles can be formed (see, \eg\ Ref.~\cite{Harrison:2009bz}), they are not so intimately related to the experimental programme.
Illustrations showing these constraints, particularly those produced by the CKMfitter~\cite{Charles:2004jd} and UTfit~\cite{Bona:2005vz} collaborations shown in Fig.~\ref{fig:ckmfit}, have become icons of the field.

The golden channel measurement of $\sin2\beta$ from $\Bd \to \jpsi \KS$ decays provides one of the strongest constraints, giving $\beta = (21.5\,^{+0.8}_{-0.7})^\circ$~\cite{Amhis:2012bh}.
BaBar and Belle have also obtained complementary measurements of $\beta$ using decays involving the quark-level transitions $b \to c\bar{c}d$ and $b \to q\bar{q}s$.
Several updated and improved measurements have become available recently~\cite{Rohrken:2012ta,Lees:2012px,Kronenbitter:2012ha,Lees:2012kxa,Lees:2011nf}, so that \CP violation effects are now established in these decay processes (see Tab.~\ref{tab:CP}).
The tensions that existed in the data some years ago have been alleviated, and all measurements of $\beta$ are consistent within their current uncertainties.

Progress on the measurement of the angle $\alpha$ has been somewhat slower in recent years, although new and improved inputs to the isospin analyses in $\pi\pi$~\cite{Lees:2012kx,Vanhoefer}, $\rho\pi$~\cite{Miyashita}, $\rho\rho$~\cite{Vanhoefer} and $a_1\pi$~\cite{Dalseno:2012hp} systems have become available.
The world average, $\alpha = (88.5\,^{+4.7}_{-4.4})^\circ$~\cite{Charles:2004jd} is dominated by the constraints from the $\rho\rho$ system, where some parameters come from only a single measurement, and others are in slight tension.  
Improved measurements, in particular of $B\to\rho\rho$ decays, are necessary to ensure that the determination is under as good control as suggested by the uncertainties.

The precision on the third angle of the Unitarity Triangle, $\gamma$, determined from $B \to DK$ and related decay processes, has stubbornly refused to go below the $10^\circ$ level despite great efforts from BaBar and Belle~\cite{delAmoSanchez:2010rq,Poluektov:2010wz}.
The latest combinations from each experiment give (BaBar) $\gamma = (69\,^{+17}_{-16})^\circ$~\cite{:2013zd,Derkach} and (Belle) $\gamma = (68\,^{+15}_{-14})^\circ$~\cite{Trabelsi}.
However, with precision measurements of some of the relevant observables now available (enabling the first $5\sigma$ observation of \CP violation in $B \to DK$ decays~\cite{Aaij:2012kz}), it appears that this barrier may soon be broken.
Indeed, the LHCb collaboration, combining results from $B \to DK$ processes~\cite{Aaij:2012kz,Aaij:2012hu,LHCb-CONF-2012-030} obtains $\gamma = (71\,^{+17}_{-16})^\circ$~\cite{John,LHCb-CONF-2012-032}.
Further improvement can be expected as more data and more decay modes are added into the combination.

The use of $B \to DK$ decays provides a theoretically clean determination of $\gamma$, since only tree-level diagrams are involved.
As such it can be considered a SM benchmark.
Measurements of $\gamma$ from charmless $B$ decays, however, can be affected by extensions of the SM due to the loop diagrams involved, but tend to suffer from larger hadronic uncertainties.
A perennial question for CKM workshops is therefore how to extract clean weak phase information from charmless hadronic $B$ decays, without losing the ability to probe beyond the SM.
Several new ideas have appeared recently in the literature~\cite{ReyLeLorier:2011ww,Ciuchini:2012gd}.

Experimentally, new data on direct \CP violation in $B$ (including \Bs) decays to $K\pi$ are available~\cite{Aaij:2012qe,Lees:2012kx,CDFnote10726,Duh:2012ie},
and the ``$K\pi$'' puzzle, namely the difference between the amount of direct \CP violation in $B$ decays to $\Km\pip$ and $\Km\piz$, persists.
Improved measurements of the $\Bz \to \KS \piz$ \CP violation parameters are necessary to test the isospin sum-rules that provide clean tests of the SM~\cite{Gronau:2005kz}.
While these must wait for data from Super $B$ factories to become available, progress on U-spin and flavour SU(3) based tests of the SM using $\Bs \to \Kp\Km$ can be expected with promising first results from LHCb already available~\cite{LHCb-CONF-2012-007}.

\begin{figure}[!htb]
\centering
\includegraphics[width=0.56\textwidth]{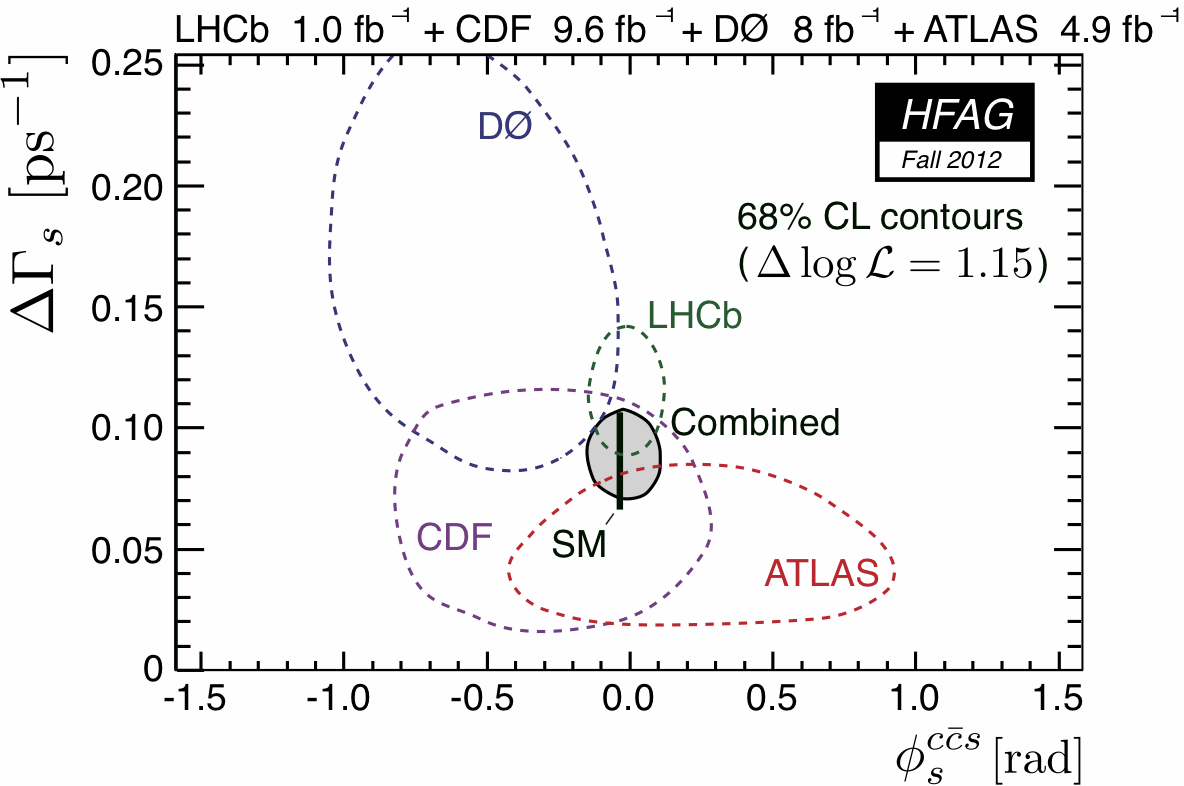}
\hspace{0.5mm}
\includegraphics[width=0.42\textwidth]{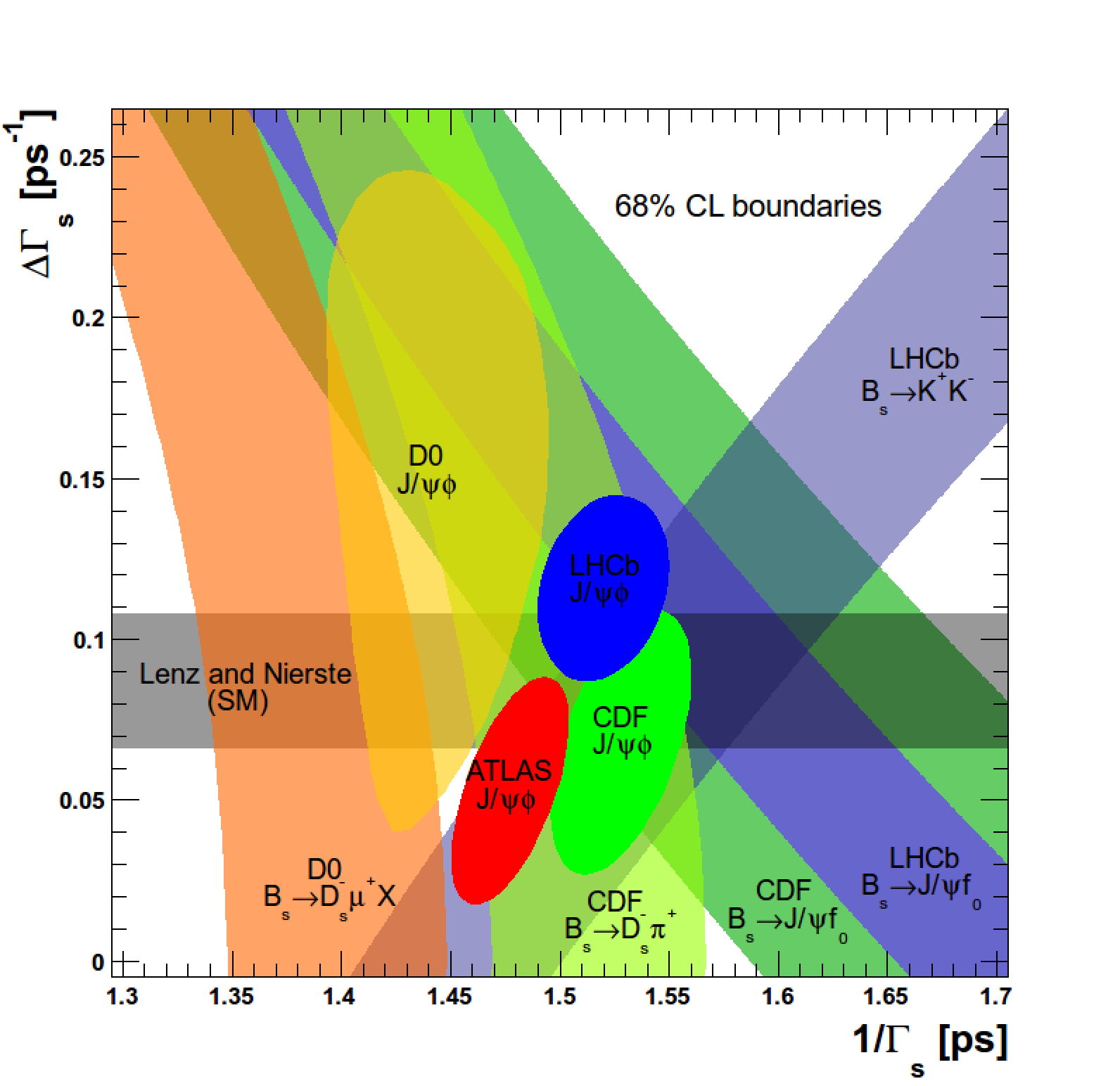}
\caption{\small
  (Left) Summary of results in the $\Delta \Gamma_s$--$\phi_s^{c \bar{c}s}$ plane~\cite{Amhis:2012bh}.
  Note that $\phi_s^{c\bar{c}s} \approx -2\beta_s$.
  (Right) Compilation of results in the $\Delta \Gamma_s$--$1/\Gamma_s$ plane~\cite{:2012fu}.
  In addition to the constraints from $\Bs \to \jpsi \phi$~\cite{Aaltonen:2012ie,Abazov:2011ry,:2012fu,LHCb-CONF-2012-002,LHCb:2012ad}, results of effective lifetime measurements in $\Bs \to \jpsi f_0(980)$~\cite{Aaltonen:2011nk,Aaij:2012nta}, $\Kp\Km$~\cite{Aaij:2012ns}, $\Dsm\pip$~\cite{Aaltonen:2011qsa} and $\Dsm\mup X$~\cite{Abazov:2006cb} are shown, and compared to the SM prediction for $\Delta \Gamma_s$~\cite{Lenz:2011ti}.
}
\label{fig:betas}
\end{figure}

As mentioned above, other triangles can be constructed based on the unitarity of the CKM matrix.
One of them is particularly relevant to the \Bs sector, and its angle $\beta_s$ (in analogy to $\beta$) can be determined using $b \to c\bar{c}s$ transitions.
Measurements of $\beta_s$ are now available from CDF~\cite{Aaltonen:2012ie}, D0~\cite{Abazov:2011ry} and ATLAS~\cite{:2012fu}\footnote{
  In contrast to the other analyses, the ATLAS measurement does not use flavour tagging.
} using $\Bs \to \jpsi \phi$ decays and from LHCb using both $\Bs \to \jpsi \phi$~\cite{LHCb:2011aa,LHCb-CONF-2012-002} and $\jpsi \pip\pim$~\cite{LHCb:2012ad} decays.
These results, summarised in Fig.~\ref{fig:betas}~(left), in particular those from LHCb, provide a significant improvement in precision (and have established $\Delta \Gamma_s > 0$~\cite{Aaij:2012eq}; a compilation of results in the $\Delta \Gamma_s$--$1/\Gamma_s$ plane is shown in Fig.~\ref{fig:betas}~(right)), while not confirming the hints of a large anomalous effect that were present in previous measurements.

The lengths of the sides of the Unitarity Triangle are determined from measurements of the magnitudes of CKM matrix elements: $\left| V_{cb} \right|$ and $\left| V_{ub} \right|$ through tree-level decays; $\left| V_{td} \right|$ and $\left| V_{ts} \right|$ through loop-dominated processes.
The latter have to date been mainly constrained from the neutral $B$ meson oscillation frequencies $\Delta m_d$ and $\Delta m_s$~\cite{LHCb-CONF-2011-050,Wishahi,Aaij:2012nt}, but loop decays such as $B \to \rho \gamma$ and $B \to \pi \mumu$ can also be used to probe $\left| V_{td} \right|$ (with corresponding $b \to s$ transitions probing $\left| V_{ts} \right|$).
Since the $\Bp \to \pip\mumu$ decay has now been observed~\cite{LHCb:2012de}, further tests of the SM using these processes become feasible.

\begin{figure}[!htb]
\centering
\includegraphics[width=0.45\textwidth,viewport=230 215 450 405,clip=true]{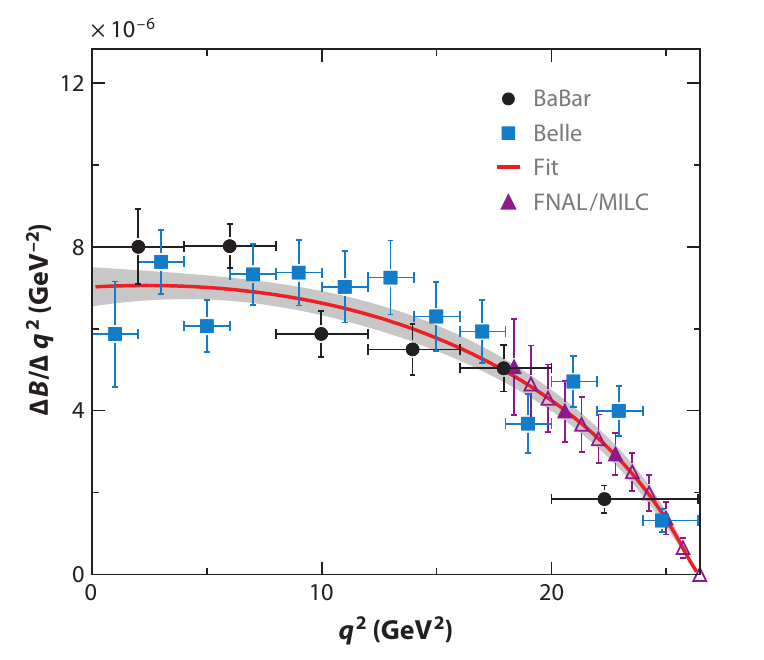}
\caption{\small
  Differential branching fraction of $\Bz \to \pim l^+ \nu$ ($l = e,\mu$)~\cite{Luth:2011zz}, using data from BaBar~\cite{delAmoSanchez:2010af} and Belle~\cite{Ha:2010rf}, and lattice QCD calculations from FNAL/MILC~\cite{Bailey:2008wp}.
}
\label{fig:pilnu}
\end{figure}

Constraints on $\left| V_{ub} \right|$ are obtained from both inclusive and exclusive processes, which have been in some tension for several years (as indeed have results on $\left| V_{cb} \right|$).
Among the exclusive modes, the $\pi l \nu$ channel has been studied in detail using a variety of experimental techniques~\cite{Lees:2012vv,delAmoSanchez:2010zd,delAmoSanchez:2010af,Ha:2010rf,Belle-LLWI2012}; a recent compilation is shown in Fig.~\ref{fig:pilnu}.

These results have recently been complemented by new results on $B \to \rho l\nu$, $\omega l\nu$, $\eta l\nu$ and $\eta^\prime l\nu$~\cite{Lees:2012vv,Lees:2012mq,delAmoSanchez:2010zd,delAmoSanchez:2010af,Belle-LLWI2012}.
The latest constraints on $\left| V_{ub} \right|$, combining BaBar and Belle data, as part of the ``Physics of the $B$ factories legacy book'' project (following the procedures developed in Ref.~\cite{Amhis:2012bh})
\begin{eqnarray*}
  \left| V_{ub} \right|_{\rm excl.} & = & \left[ 3.23 \times \left( 1.00 \pm 0.05_{\rm exp} \pm 0.08_{\rm th} \right) \right] \times 10^{-3} \, .\\
  \left| V_{ub} \right|_{\rm incl.} & = & \left[ 4.42 \times \left( 1.000 \pm 0.045_{\rm exp} \pm 0.034_{\rm th} \right) \right] \times 10^{-3} \, .
\end{eqnarray*} 
An average of these numbers has a $\chi^2$ (probability) of 3.0 (0.003).
The origin of the discrepancy is not understood, and so the uncertainty is scaled by $\sqrt{3}$ to give 
\begin{eqnarray*}
  \left| V_{ub} \right|_{\rm avg.} & = & \left[ 3.95 \times \left( 1.000 \pm 0.096_{\rm exp} \pm 0.099_{\rm th} \right) \right] \times 10^{-3} \, .
\end{eqnarray*} 
Better understanding will be needed to reduce the uncertainty.

A complementary determination of $\left| V_{ub} \right|$ can be made from the leptonic decay $\Bp \to \taup\nu$, using as input the lattice QCD calculation of the \Bp decay constant.  
Such a measurement is sensitive to extensions to the SM, such as models containing charged Higgs bosons.
Previous results had shown a deviation with the prediction based on the CKM fit, but the latest result from Belle~\cite{Adachi:2012mm} reduces the tension to below the $2\sigma$ level.
It should be noted that the signals for $\Bp \to \taup\nu$~\cite{Aubert:2009wt,Lees:2012ju,Hara:2010dk,Adachi:2012mm} are individually below the $5\sigma$ threshold usually demanded for discovery, further indicating that claims of a non-SM excess were premature.

Another channel that is sensitive to potential effects of charged Higgs bosons is $B \to D^{(*)}\tau\nu$.
These four decays (\Bp or \Bz to final states containing $D$ or $D^*$) are all now observed with above $5\sigma$ significance~\cite{Bozek:2010xy,Lees:2012xj}, with the latest results giving an excess above the SM prediction at $3.4\sigma$ significance~\cite{Lees:2012xj}.\footnote{
  The Belle results~\cite{Bozek:2010xy} are not included in this combination since correlations that are not publicly available need to be taken into account.
}
Moreover the pattern of the excess in $D\tau\nu$ and $D^*\tau\nu$ is inconsistent with that expected in the simplest model containing charged Higgs bosons, the (type II) two-Higgs doublet model~\cite{Tanaka:2010se}.
This demands further experimental investigation.

\section{Rare decays}

The two most common generic arguments in favour of a continued experimental programme in quark flavour physics are (i) there must be new sources of \CP violation that may manifest themselves in the quark sector (through measurements such as those discussed above); (ii) precise measurements of ``rare decays'', \ie\ processes that are suppressed in the SM, are sensitive to physics beyond the SM at scales that can exceed those that can studied at the energy frontier.
Both of these remain strong arguments, although it must be recognised that neither constitutes a guarantee of discovery in any particular experiment.
In addition, the recent measurement of a large value of the lepton mixing angle $\theta_{13}$~\cite{An:2012eh,Ahn:2012nd} suggests that the observation of \CP violation in neutrino oscillations may become a reality within the next decade or so.  
This provides a ``best before'' date for the \CP violation-based argument.
On the other hand, the observation of a Higgs-like particle by ATLAS~\cite{:2012gk} and CMS~\cite{:2012gu}, and the non-discovery of any non-SM particle, somewhat strengthens the argument for precision probes of rare processes (including \CP-violating processes) that are sensitive to physics at high scales.

The archetypal flavour-changing neutral-current decay is the $b \to s\gamma$ transition.  
The rate has been measured to be consistent with the SM prediction~\cite{Misiak:2006zs} within $\sim 10\,\%$ uncertainty, but further studies remain worthwhile, and new results are becoming available for both inclusive~\cite{Lees:2012ym,:2012iwb} and exclusive decays~\cite{Aaij:2012st}.
In particular, improved measurements of \CP asymmetries and of the polarisation of the emitted photon will help to constrain, model-independently, the phase space of extensions to the SM.

\begin{figure}[!htb]
\centering
\includegraphics[width=0.45\textwidth]{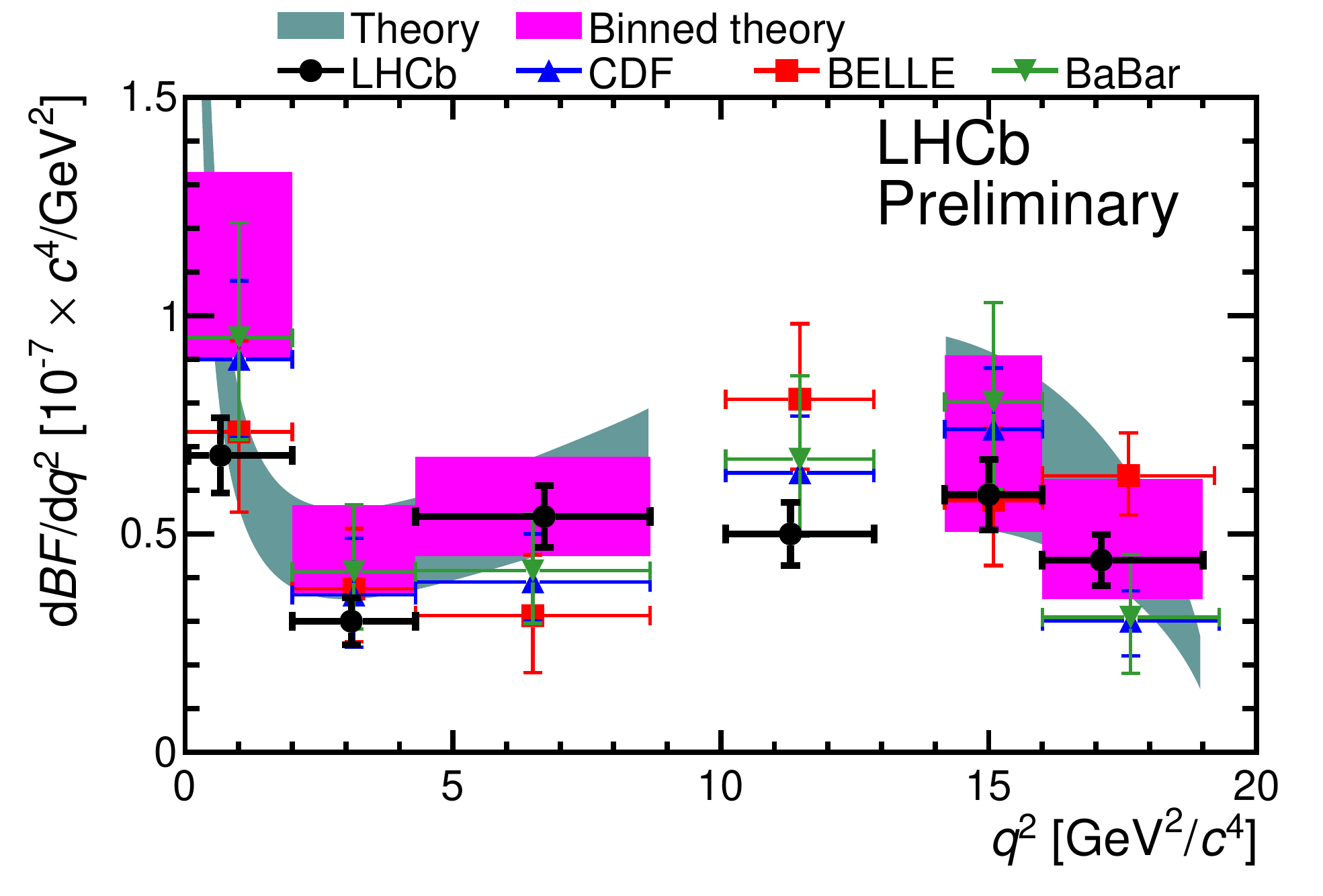}
\hspace{3mm}
\includegraphics[width=0.45\textwidth]{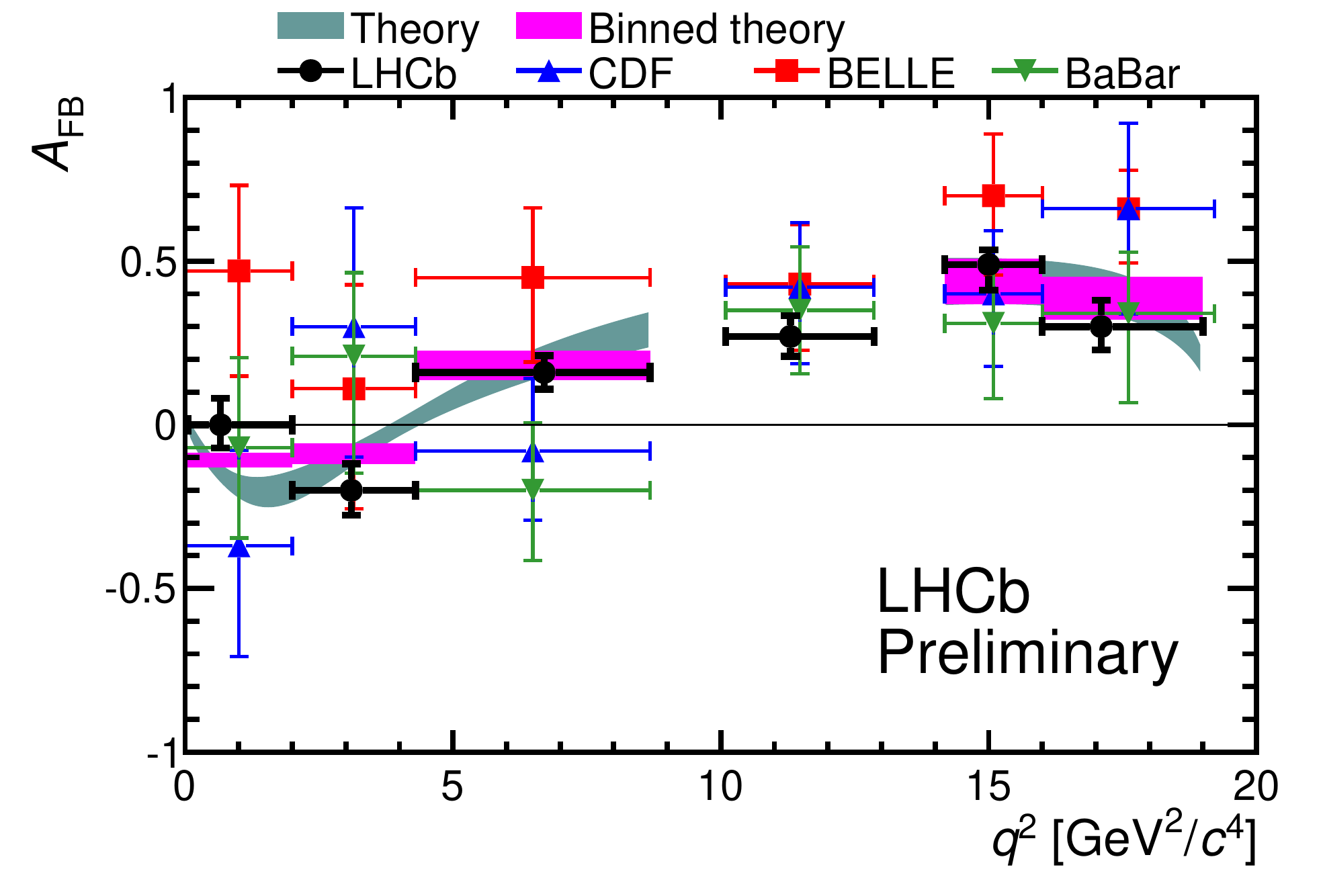} \\
\includegraphics[width=0.45\textwidth]{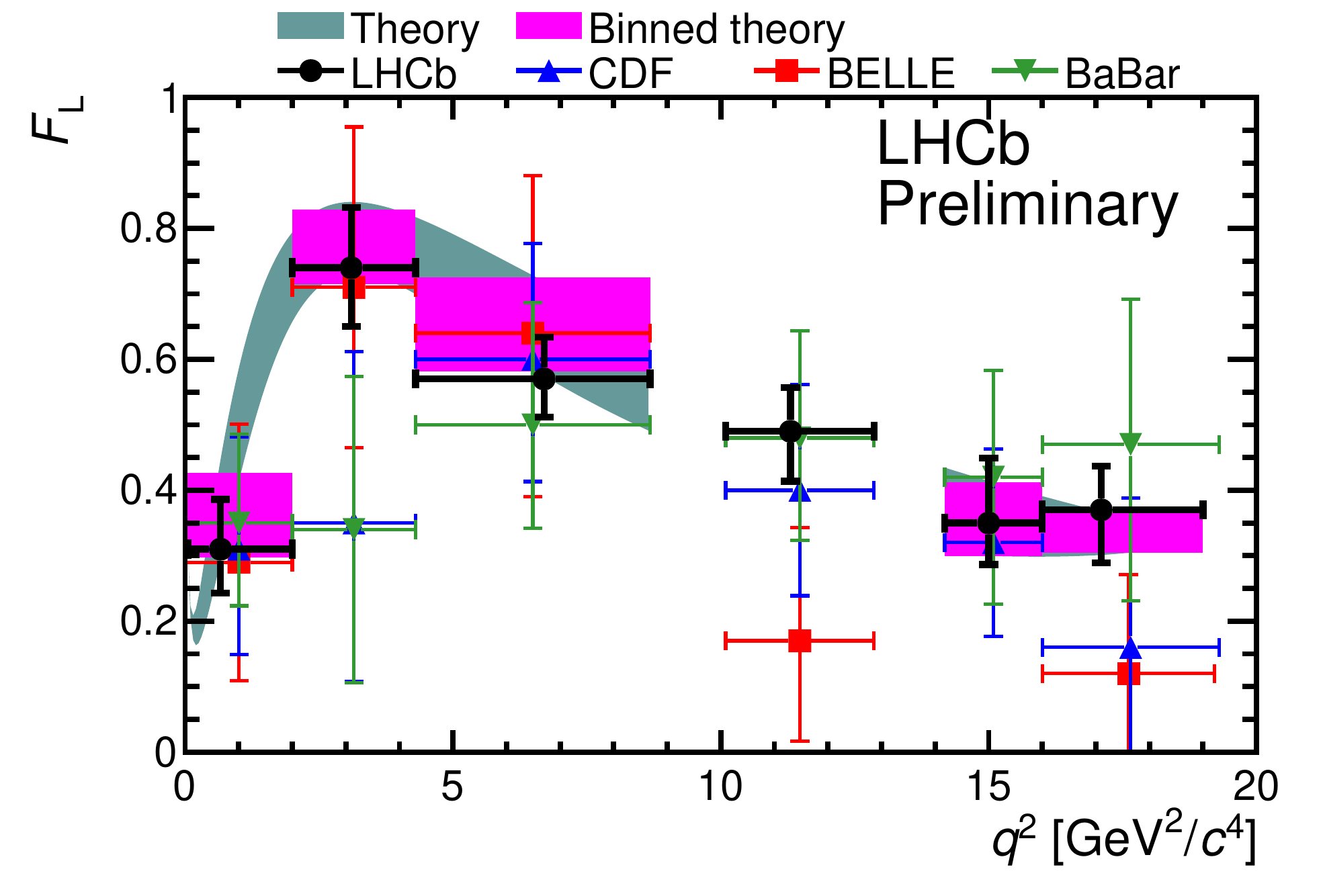}
\hspace{3mm}
\includegraphics[width=0.45\textwidth]{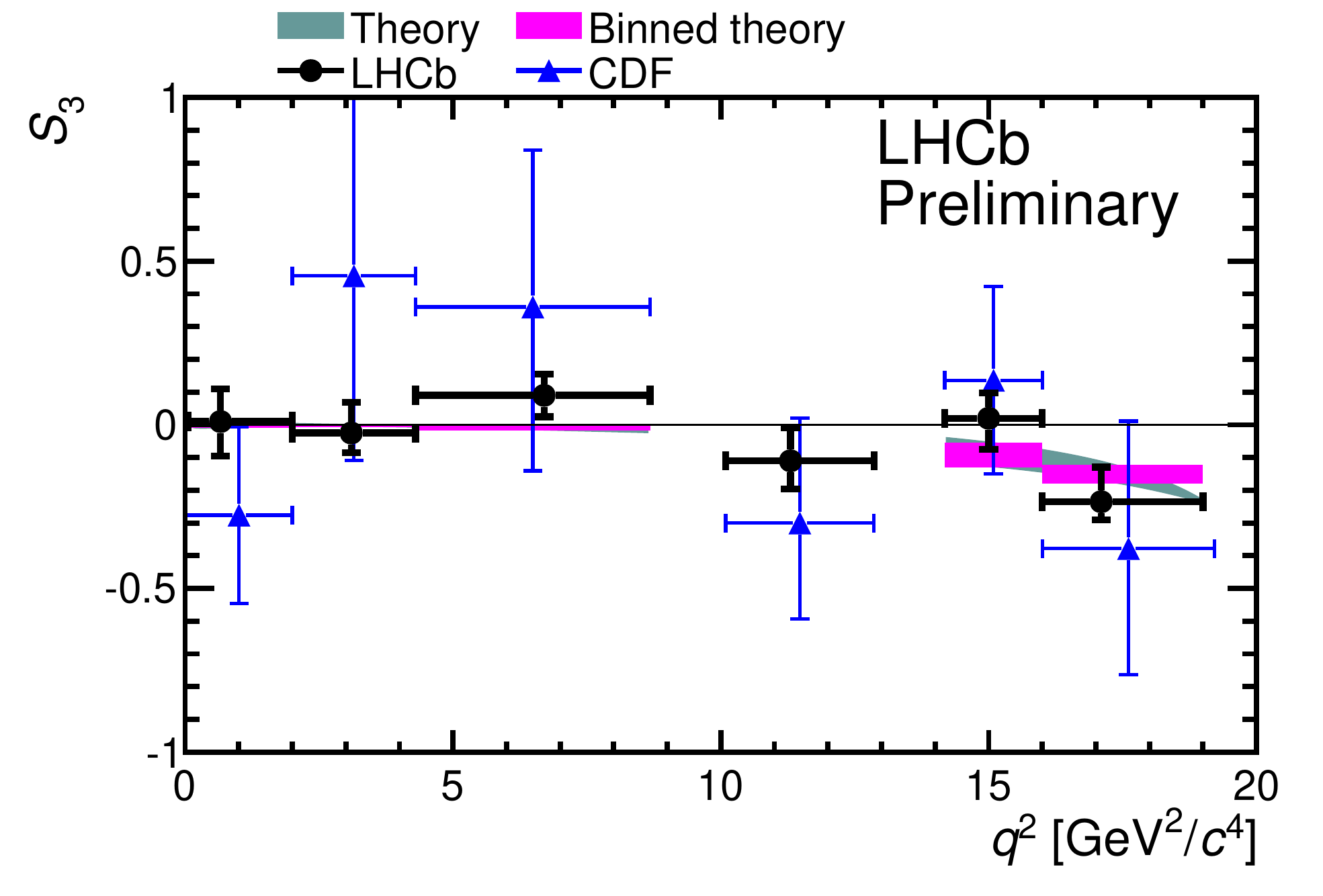}
\caption{\small
  Compilation of measurements of (top left) the differential branching fraction, (top right) the forward-backward asymmetry, (bottom left) the longitudinal polarisation fraction, (bottom right) the parameter $S_3$ in $\Bd\to \Kstarz\mumu$ decays (from Ref.~\cite{LHCb-CONF-2012-008}), as functions of the dimuon invariant mass squared.
  The data points are compared to the SM expectations~\cite{Bobeth:2011gi}.
}
\label{fig:kstarmumu}
\end{figure}

The $B \to K^*\mumu$ decay provides an excellent laboratory to test the SM, via precise measurements of the Wilson coefficients in the operator product expansion of the effective weak Hamiltonian.  
Recently new results have become available from BaBar~\cite{Ritchie}, CDF~\cite{Behari} and LHCb~\cite{Aaij:2011aa,LHCb-CONF-2012-008}.
The latest LHCb analysis has the smallest uncertainties, and allows for the first time a determination of the zero-crossing point of the forward-backward asymmetry, $q^2_0 = (4.9\,^{+1.1}_{-1.3}) \gev^2/c^4$, consistent with the SM prediction~\cite{Bobeth:2011nj,Beneke:2004dp,Ali:2006ew}.
A compilation, including also results from Belle~\cite{Wei:2009zv} is shown in Fig.~\ref{fig:kstarmumu}.
However, there is still a long road to the final goal of a full angular analysis of $B \to K^*\mumu$ decays.

The highest profile results in heavy flavour physics must certainly be those on the $\Bs \to \mumu$ decay.
This decay is suppressed in the SM due to three effects: the absence of tree-level flavour-changing-neutral currents, the CKM factors, and the V-A structure of the weak interaction (helicity suppression).
Many extensions to the SM break at least one of these factors -- of particular interest is that models with extended Higgs' sectors (such as SUSY) may break the helicity suppression.
This can potentially lead to large enhancements of the branching fraction over the SM prediction~\cite{Buras:2012ru}, which is precisely known, in part due to the good control over the lattice QCD calculations which are used as inputs (as reviewed, \eg\ in Ref.~\cite{Bediaga:2012py}).

\begin{figure}[!htb]
\centering
\includegraphics[width=0.43\textwidth,viewport=0 500 384 735,clip=true]{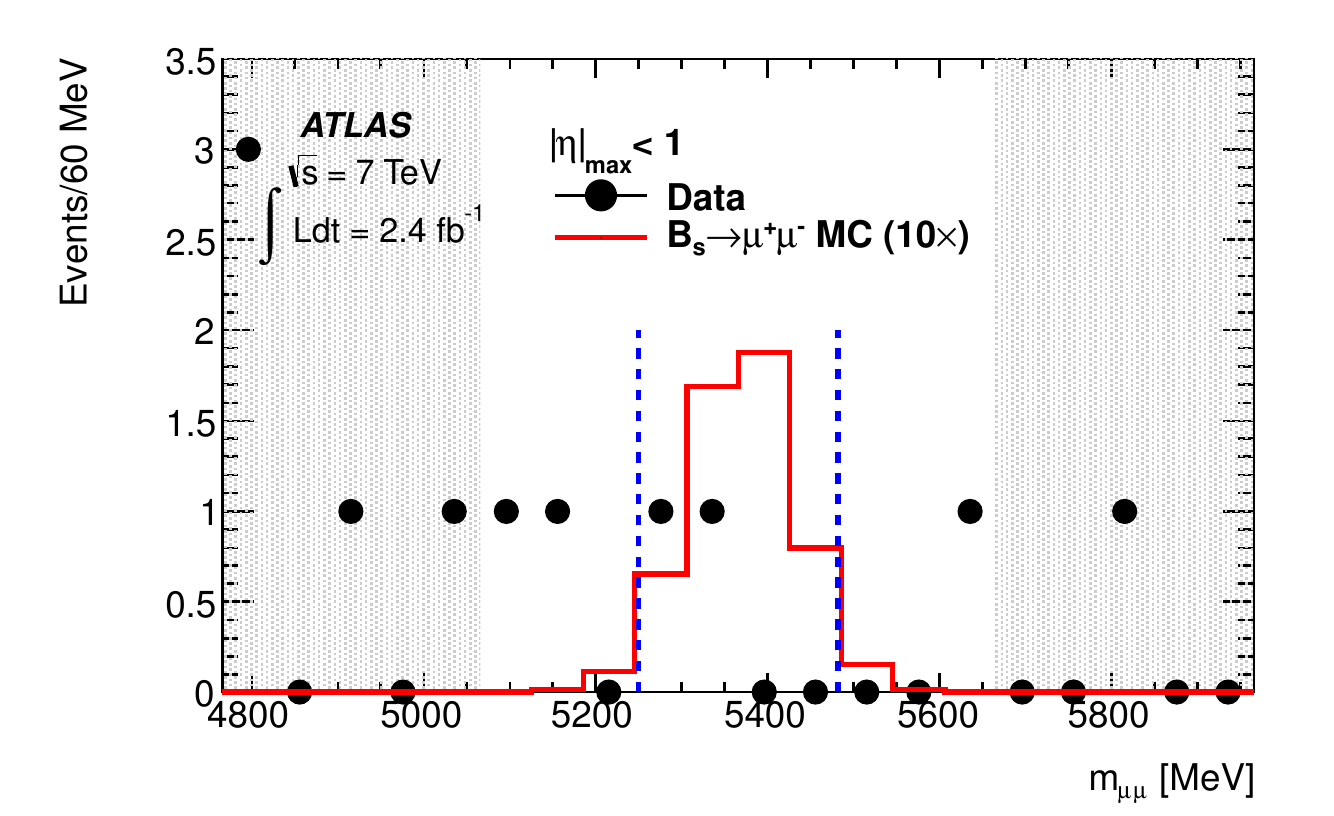}
\hspace{1mm}
\includegraphics[width=0.27\textwidth]{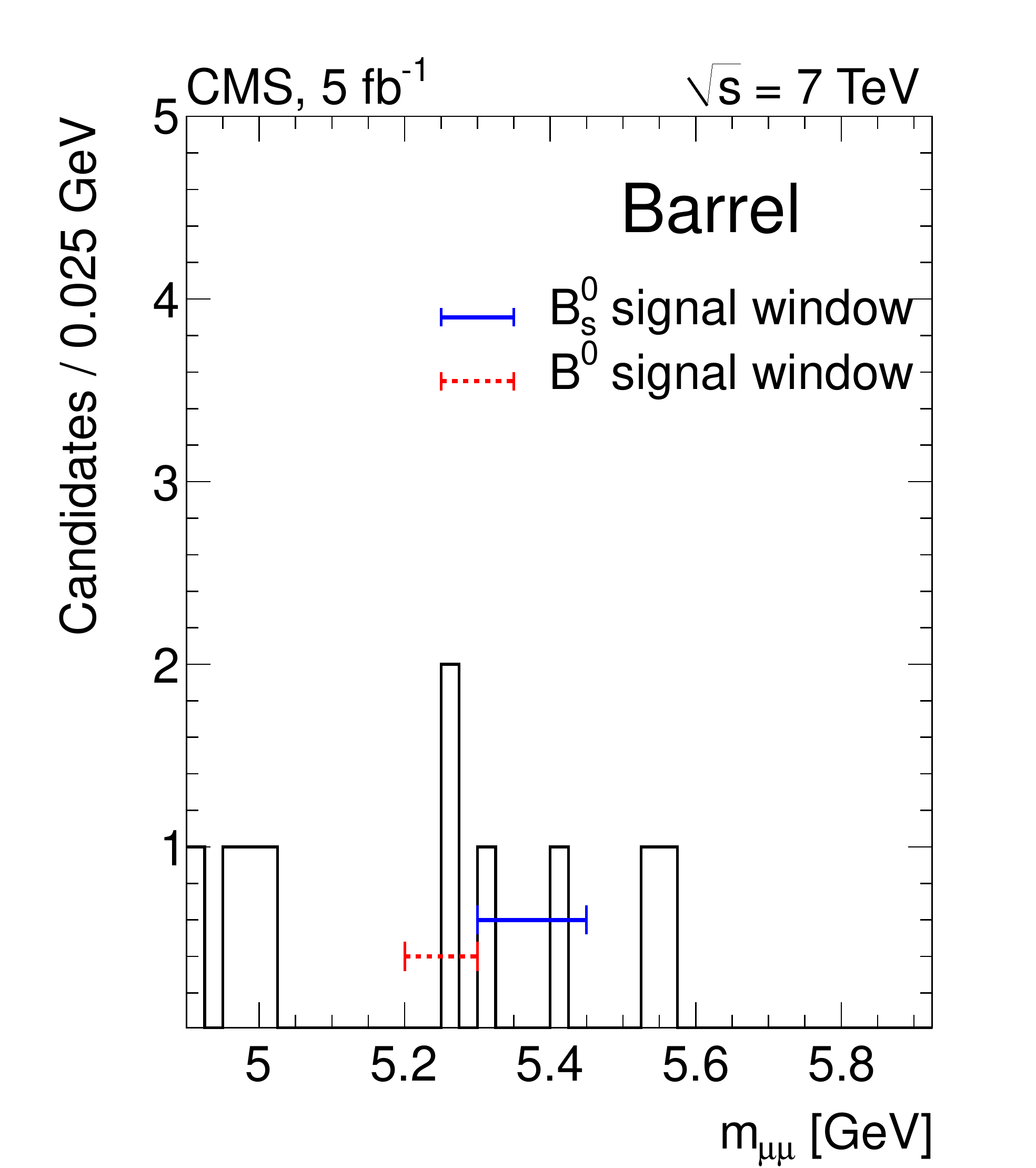}
\hspace{1mm}
\includegraphics[width=0.25\textwidth]{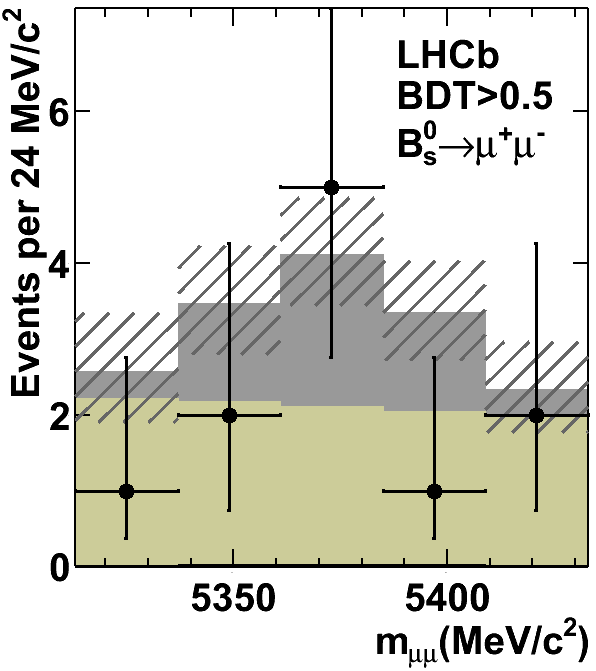}
\caption{\small
  Latest results on the search for $\Bs\to\mumu$ from (left) ATLAS~\cite{Aad:2012pn}, (middle) CMS~\cite{Chatrchyan:2012rga}, (right) LHCb~\cite{Aaij:2012ac}.
}
\label{fig:bsmumu}
\end{figure}

The latest results from ATLAS, with $2.4 \invfb$~\cite{Aad:2012pn}, CMS, with $5.0 \invfb$~\cite{Chatrchyan:2012rga} and LHCb, with $1.0 \invfb$~\cite{Aaij:2012ac} (all using $7 \tev$ LHC $pp$ collision data recorded in 2011) are shown in Fig.~\ref{fig:bsmumu}.
The most restrictive limit from a single experiment is the LHCb result ${\cal B}(\Bs\to\mumu)<4.5 \times 10^{-9}$ at 95\,\% confidence level~\cite{Aaij:2012ac}, while a combination of LHC results gives ${\cal B}(\Bs\to\mumu)<4.2 \times 10^{-9}$ at 95\,\% confidence level~\cite{LHCb-CONF-2012-017}.
The results approach the sensitivity required to probe the SM expectation, and updates including the 2012 LHC data are hotly anticipated.\footnote{
  While these proceedings were in preparation, the LHCb collaboration announced the first evidence for the $\Bs \to \mumu$ decay, with branching fraction ${\cal B}(\Bs \to \mumu) =  (3.2\,^{+1.5}_{-1.2}) \times 10^{-9}$~\cite{:2012ct}.
}

\section{``Bread and butter'' measurements}

In any summary of a wide range of experimental results, it is natural to focus attention mainly on ``golden channels''.
However, by doing so it is possible that the importance of making improved measurements of other, less glamorous, observables can be overlooked.
It should be stressed that such measurements are an integral part of the overall programme.
Indeed, very often the most highly cited results depend directly on input from other measurements.

An excellent example comes from the $\Bs \to \mumu$ analysis: in order to measure the absolute branching fraction of any \Bs decay, it is necessary to know the relative production rates of \Bs and \Bd mesons, denoted as $f_s/f_d$. 
Then the branching fraction can be measured relative to that of a well-known \Bd (or \Bu, isospin symmetry is assumed) decay -- typically $\Bp \to \jpsi \Kp$ is used .
Measurement of $f_s/f_d$~\cite{Aaij:2011hi,Aaij:2011jp}, however, relies on knowledge of absolute branching fractions of charmed hadrons, in particular of ${\cal B}(\Dsp \to \Kp\Km\pip)$.
This value is currently known to $5.1\,\%$ precision~\cite{Alexander:2008aa}, and future measurements of ${\cal B}(\Bs \to\mumu)$ will be limited at this level unless improved results become available.
Therefore it is very encouraging that a new preliminary result on ${\cal B}(\Dsp \to \Kp\Km\pip)$ is available from Belle (as a spin-off from the measurement of ${\cal B}(\Dsp \to\taup\nu)$~\cite{Belle-Charm2012}).

This is just one example of the large impact that ``bread and butter'' measurements can have.  
Other examples include lifetime and mass measurements.
It should also be remembered that discoveries may appear in unexpected places.

\section{Summary}

Sometime Cincinnati resident Mark Twain once wrote, ``{\it The most permanent lessons in morals are those which come, not of booky teaching, but of experience.}''
This can be interpreted as encouragement for the experimentalist.
To summarise, then, the following (hopefully) permanent lessons have been learned:
\begin{itemize}
\item The foundation of accelerator-based physics is, of course, excellent performance by the accelerators.
\item Investment in detectors and techniques brings rewards.
\item Interesting effects could be big ...
  \begin{itemize}
  \item but they may be very small, so be prepared to be precise,
  \item it seems like there are no ${\cal O}(1)$ deviations from the SM.
  \end{itemize}
\item
  Clean theoretical predictions are to be treasured ...
  \begin{itemize}
  \item but in their absence, data-driven methods to control uncertainties are to be valued.
  \end{itemize}
\item $3\sigma$ effects seem to often go away, but $5\sigma$ effects seem to stay ...
  \begin{itemize}
  \item nevertheless, investigating anomalies is worth the effort, and is sure to result in something being learned (whether about physics, systematics or statistics).
  \end{itemize}
\item Bread and butter measurements provide an essential appetiser to the feast of interesting observables that heavy flavour experiments can study.
\item New physics just might be around the corner ...
  \begin{itemize}
  \item so there is plenty to look forward to in heavy quark physics in the coming years~\cite{Roney,Schune}.
  \end{itemize}
\end{itemize}
Whether or not these lessons have taught us any morals must be left to CKM2014.

\section*{Acknowledgements}

\noindent The author wishes to thank the organisers for the invitation, and to congratulate them for a productive and enjoyable workshop.
Vincenzo Chiochia, Rob Harr, Mikihiko Nakao, Abi Soffer and Mark Williams provided suggestions for the content of the talk.
Rick van Kooten, Marie-H\'{e}l\`{e}ne Schune and Karim Trabelsi commented on drafts of the proceedings.
Errors and omissions are the sole fault of the author.
This work is supported by the 
Science and Technology Facilities Council (United Kingdom), CERN, 
and by the European Research Council under FP7.

\bibliographystyle{LHCb}
\bibliography{main,LHCb-CONF}

\end{document}